\documentstyle{cupconf}
\newcommand{\apgt}{\ {\raise-.5ex\hbox{$\buildrel>\over\sim$}}\ }
\newcommand{\aplt}{\ {\raise-.5ex\hbox{$\buildrel<\over\sim$}}\ }
\begin{document}
\title[The Progenitors of Type Ia Supernovae]
{The Progenitors of Type Ia Supernovae}
\author[M.\ Livio]{M\ls A\ls R\ls I\ls O\ns L\ls I\ls V\ls I\ls O}
\affiliation{Space Telescope Science Institute, 3700 San Martin Drive,
Baltimore, MD 21218}
\maketitle
\begin{abstract} 
Models for Type~Ia Supernovae (SNe~Ia) are reviewed. It is shown that
there are strong reasons to believe that SNe~Ia represent thermonuclear
disruptions of C--O white dwarfs, when these white dwarfs reach the
Chandrasekhar limit and ignite carbon at their centers.

Progenitor scenarios are reviewed critically and the strengths and
weaknesses of each scenario are explicitly presented. It is argued that
single-degenerate models, in which the white dwarf accretes from a
subgiant or giant companion are currently favored. The relation of the
different models to the use of SNe~Ia for the determination of
cosmological parameters is also discussed.

Observational tests of the conclusions are suggested.
\end{abstract}
\section{Introduction}
During the past year two groups have presented strong evidence that the
expansion of the universe is accelerating rather than decelerating
(Perlmutter et~al.\ 1998, 1999; Schmidt et~al.\ 1998; Riess et~al.\
1998). This surprising result comes from distance measurements to about
fifty supernovae Type~Ia in the redshift range $z=0$ to $z=1$. The
results are consistent with the cosmological constant (or vacuum energy)
contributing to the total energy density about 70\% of the critical
density.

This unexpected finding, as well as the use of supernovae Type~Ia to
measure the Hubble constant (e.g.\ Sandage et~al.\ 1996; Saha et~al.\
1997), has focused the attention again on the
frustrating fact that in spite of decades of research the exact nature
of the progenitors of supernovae Type~Ia remains unknown. Until this
problem is solved, one cannot be fully confident that supernovae at
higher redshifts are not somehow different from their low redshift
counterparts. In the present review I therefore examine critically the
question of the nature of the progenitors of supernovae Type~Ia. Other
recent reviews include Branch et~al.\ (1995), Livio (1996a), Renzini
(1996)
and Iben (1997).

\section{Characteristics and the basic model}

The {\it defining\/} characteristics of supernovae Type~Ia (SNe~Ia) are
both spectral: (i)~the {\it lack\/} of lines of hydrogen, and (ii)~the
{\it
presence\/} of a strong red Si~II absorption feature ($\lambda$6355
shifted to $\sim6100$~\AA).

Once defined as SNe~Ia, the following are several of the important {\it
observational characteristics\/} of the class which may help in the
search for progenitors:
\begin{enumerate}
\item[(1)] {\it Homogeneity\/}: Nearly 90\% of all SNe~Ia form a
homogeneous
class in terms of their {\it spectra\/}, {\it light curves\/}, and {\it
peak absolute magnitudes\/}. The latter are given by
\begin{equation}
M_{\rm B}\simeq M_{\rm V}\simeq-19.30(\pm0.03)+5\log (H_0/60~{\rm
km~s}^{-1}~{\rm
Mpc}^{-1})
\end{equation}
with a dispersion of $\sigma(M_{\rm B})\sim\sigma(M_{\rm
V})\sim0.2$--0.3
(Hamuy et~al.\ 1996a; Tamman \& Sandage 1995; and see Branch 1998 for a
review).
\item[(2)] {\it Inhomogeneity\/}: Some differences in the spectra and
light
curves do exist (e.g.\ Hamuy et~al.\ 1996b). In terms of explosion
strength, SNe~Ia can roughly be ordered as follows: SN~1991bg and
SN~1992K
represent the weakest events, followed by weak events like 1986G,
followed by about 90\% of all SNe~Ia which are called ``normals'' (or
``Branch normals''), to the stronger than normal events like SN~1991T.
\item[(3)] The {\it luminosity function\/} of SNe~Ia declines very
steeply on
the bright side (e.g.\ Vaughan et~al.\ 1995). Since selection effects
cannot prevent the discovery of SNe which are brighter than the
``normals,'' this implies that {\it the normals are essentially the
brightest\/}.
\item[(4)] Near maximum light, the spectra are characterized by {\it
high
velocity\/} (8000--30,000 km~s$^{-1}$) {\it intermediate mass
elements\/} (O--Ca). In the late, nebular phase, the spectra are
dominated by forbidden lines of iron (e.g.\ Kirshner et~al.\ 1993;
Wheeler et~al.\ 1995; Ruiz-Lapuente et~al.\ 1995; G\'omez et~al.\ 1996;
Filippenko 1997).
\item[(5)] Fairly young populations appear to be most efficient at
producing
SNe~Ia (e.g.\ they tend to be associated with spiral arms in spirals;
Della Valle \& Livio 1994; Bartunov, Tsvetkov \& Filimonova 1994), but
relatively old populations ($\tau\apgt4\times10^9$~yr) can also produce
them. In particular, {\it SNe~Ia do occur in ellipticals\/} (e.g.\
Turatto, Cappellaro \& Benetti 1994). This immediately implies that
{\it SNe~Ia are not caused by the core collapse of stars more massive
than 8~M$_{\odot}$\/}.
\item[(6)] There exist a number of correlations between different pairs
of
observables (see e.g.\  Branch 1998 for a review). Of these, the most
frequently used in the context of determinations of cosmological
parameters is the correlation between the {\it absolute magnitude and
the
shape of the light curve\/}. Basically, brighter SNe~Ia decline more
slowly. The parameter commonly used to quantify the light curve shape is
$\Delta m_{15}$ (Phillips 1993), the decline in magnitudes in the
$B$~band during the first 15~days after maximum light. Hamuy et~al.\
(1996a) find slopes $dM_{\rm B}/d\Delta m_{15}=0.78\pm 0.17$, $dM_{\rm
V}/d\Delta m_{15}=0.71\pm 0.14$, and $dM_{\rm I}/d\Delta m_{15}=0.58\pm
0.13$. Using a stretch-factor $s$ (Perlmutter et~al.\ 1997), one can
write $M_{\rm B}=M_{\rm B}(s=1)-\alpha *(s-1)$, with $M_{\rm B}
(s=1)=-19.46$ (e.g.\ Sandage et~al.\ 1996), and $\alpha=1.74$
(Perlmutter et~al.\ 1999). Sophisticated techniques for using the
different 
correlations in distance determinations have been developed (e.g.\ Riess 
et~al.\ 1996, 1998).
\end{enumerate}

The above characteristics can be augmented with the following suggestive
facts:
\begin{enumerate}
\item[(1)] The {\it energy\/} per unit mass,
$1/2(\sim10^4$~km~s$^{-1})^2$, is
of the order of the one obtained from the conversion of carbon and
oxygen to iron.
\item[(2)] The fact that the event is explosive suggests that {\it
degeneracy\/} may play a role.
\item[(3)] The spectrum contains no hydrogen.
\item[(4)] The explosions can occur with long delays, after the
cessation of
star formation.
\end{enumerate}

All the properties above have led to one agreed upon model: {\it SNe~Ia
represent thermonuclear disruptions of mass accreting white dwarfs\/}.

It is interesting that there exists a unanimous consensus on this model
in spite of the fact that the essence of flame physics and the details
of the transition from deflagration to detonation (in particular the
density at which the transition occurs), which are at the heart of the 
model, remain as major unsolved problems (e.g.\ Khokhlov, Oran \&
Wheeler 
1997; Woosley 1997; Reinecke, Hillebrandt \& Niemeyer 1998; and talks 
by Khokhlov, Arnett, and Hillebrandt presented at the Chicago meeting on 
Type~Ia Supernovae: Theory and Cosmology, October 1998). In fact, given 
these uncertainties in the deflagration to detonation transition it is 
almost difficult to understand how the entire family of SNe~Ia light 
curves can be fitted essentially with one parameter (e.g.\ Perlmutter 
et~al.\ 1997), although it is possible that all SNe~Ia explode at the 
same WD mass (see \S4), and that the entire observed diversity stems 
from different $^{56}$Ni masses.

\section{Why is identifying the progenitors important?}

The fact that we do not know yet what are the progenitor systems of some 
of the most dramatic explosions in the universe has become a major
embarrassment and one of the key unsolved problems in stellar evolution.
There are several important reasons why identifying the progenitors has
become more crucial than ever: 
\begin{enumerate}
\item[(i)] The use of SNe~Ia as one of the main ways to determine the
key cosmological parameters $H_0$, and the contribution to the energy 
density $\Omega_{\rm M}$, $\Omega_{\Lambda}$ requires an understanding 
of the evolution of the luminosity,  and the SN rate with cosmic epoch. 
Both of these depend on the nature of the progenitors.
\item[(ii)] Galaxy evolution depends on the radiative, kinetic energy,
and nucleosynthetic output of SNe~Ia (e.g.\ Kauffmann, White \&
Guiderdoni 1993).
\item[(iii)] Due to the uncertainties that still exist in the explosion
mechanism itself, a knowledge of the initial conditions and of the
distribution of matter in the environment of the exploding star are
essential for the understanding of the explosion.
\item[(iv)] An unambiguous identification of the progenitors, coupled
with observationally determined SNe~Ia rates will help to place
meaningful constraints on the theory of binary star evolution (e.g.\
Livio 1996b; Li \& van den Heuvel 1997; Yungelson \& Livio 1998;
Hachisu,
Kato \& Nomoto 1999). In particular, a semi-empirical determination of
the elusive common-envelope efficiency parameter, $\alpha_{\rm CE}$, may
be possible (e.g.\ Iben \& Livio 1993).
\end{enumerate}

\section{Refinements of the basic model}

The basic model for SNe~Ia (that essentially all researchers in the
field
agree upon) is that of a thermonuclear disruption of an accreting white
dwarf (WD). However, additional refinements to the model are possible on
the basis of existing observational data and theoretical models. These
refinements still do not involve the question of the {\it progenitor
systems\/}. Rather, they address the question of the WD {\it
composition\/}, and of its {\it mass\/} at the instant of explosion.

\subsection{The composition of the exploding WD}

In principle, the WD that accretes to the point of explosion could be
composed of He, of C--O, or of O--Ne. Let us examine these possibilities
one by one.
\begin{enumerate}
\item[(i)] {\it He WDs\/}: Helium WDs have typical masses that are
smaller 
than $\sim0.45$~M$_{\odot}$ (e.g.\ Iben \& Tutukov 1985). While if
accreting,
these He WDs can explode following central He ignition at
$\sim0.7$~M$_{\odot}$, the composition of the ejected matter in this
case will be that of He, $^{56}$Ni and decay products (e.g.\ Nomoto \&
Sugimoto
1977; Woosley, Taam \& Weaver 1986). This is entirely inconsistent with
observations (observational characteristic~(4) in \S2). Therefore, {\it
He WDs certainly do not produce the bulk of SNe~Ia\/}.
\item[(ii)] {\it O--Ne WDs\/}: Oxygen--Neon WDs form in binaries from
main sequence stars of $\sim10$~M$_{\odot}$, although the precise range
which allows formation is somewhat uncertain (e.g.\ Iben \& Tutukov
1985; Canal, Isern \& Labay 1990; Dominguez, Tornamb\'e \& Isern
1993). These systems are probably not numerous enough to constitute the
main channel of SNe~Ia (e.g.\ Livio \& Truran 1992). It is also
generally expected that O--Ne WDs that manage to accrete enough material
to reach the Chandrasekhar limit will produce preferentially 
accretion-induced collapses (to form neutron stars) rather than SNe~Ia
(e.g.\ Nomoto \& Kondo 1991; Gutierrez et~al.\ 1996). I should note
that the existing calculations have been performed for WDs of 
O--Ne--Mg composition, while recent calculations of the evolution of a
10~M$_{\odot}$ star produce degenerate cores which are almost devoid of
magnesium (Ritossa, Garcia-Berro \& Iben 1996). Nevertheless, because of
the above two points {\it it is unlikely that O--Ne WDs produce the bulk
of SNe~Ia\/}.
\item[(iii)] {\it C--O WDs\/}: Carbon--Oxygen WDs are formed in binaries
from main sequence stars of up to $\sim10$~M$_{\odot}$. They are
therefore both relatively numerous, and they provide a significant
``phase
space volume'' (masses in the range 0.8--1.2~M$_{\odot}$; accretion
rates in the range $10^{-8}$--$10^{-6}$~M$_{\odot}$/yr) in which they
are expected to produce SNe~Ia (upon reaching the Chandrasekhar limit;
e.g.\ Nomoto \& Kondo 1991). Consequently, {\it the accreting WDs that
produce most of the SNe~Ia are very probably of C--O composition\/}!
\end{enumerate}

\subsection{At what mass does the WD explode and where and in what fuel
does the ignition take place?}

While there is virtually unanimous agreement about everything I said up
to now, namely, that: {\it SNe~Ia are thermonuclear disruptions of
accreting C--O WDs\/}, the next step in the refinement of the model is
more controversial. Two major classes of models have been considered,
and they suggest entirely different answers to the questions posed by
the title of this subsection. In one class, the WD explodes upon
reaching the {\it Chandresekhar mass\/}, as {\it carbon\/} ignites at
its {\it center\/}. In the second, the WD explodes at a {\it 
sub-Chandresekhar mass\/}, as {\it helium\/} ignites {\it off-center\/}.
I will now review briefly each of these classes and point out its
strengths and weaknesses.

\subsubsection{Chandrasekhar mass carbon ignitors}

In this model, considered `standard,' the WD accretes until it
approaches the Chandrasekhar mass. Carbon ignition occurs at or very
near the center and the  burning front propagates outwards. The main
{\it strengths\/} of this model are (see e.g.\ Hoeflich \& Khokhlov
1996; Nugent et~al.\ 1997 for detailed modeling):
\begin{enumerate}
\item[(1)] Some {\it $10^{51}$~ergs of kinetic energy\/} are deposited
into the ejecta by nuclear energy.
\item[(2)] $^{56}$Ni decay powers the {\it lightcurve\/}.
\item[(3)] The density and composition as a function of the ejection of
velocity (X$_{\rm i}$(V$_{\rm ej}$)) are consistent with the observed
{\it
spectra\/}.
\item[(4)] The fact that the explosion occurs at the Chandrasekhar mass 
explains the {\it homogeneity\/}.
\item[(5)] Spectra (e.g.\ of SNe~1994D, 1992A) can be fitted in great
detail by theoretical models (e.g.\ Nugent et~al.\ 1997).
\end{enumerate}

The main {\it weaknesses\/} of the Chandrasekhar mass models are:
\begin{enumerate}
\item[(1)] It has proven more difficult than originally thought for WDs
to accrete {\it up to the Chandrasekhar mass\/} in sufficient numbers
to account for the SNe~Ia rate. The difficulty is associated with mass
loss episodes in nova explosions, in helium shell flashes and in massive
winds or
common envelope phases. I will return to some of these problems when I
discuss specific progenitor models.
\item[(2)] For initial WD masses larger than $\sim1.2$~M$_{\odot}$, {\it
accretion-induced collapse\/} is a more likely outcome than a SN~Ia
(e.g.\ Nomoto \& Kondo 1991).
\item[(3)] {\it The late-time spectrum\/} ($\sim300$ days), and in
particular the Fe~III feature at \mbox{$\sim4700$~\AA}\ does not agree
well
with Chandrasekhar mass models (Liu, Jeffrey \& Schultz 1998).
\end{enumerate}

My overall assessment of Chandrasekhar mass models is that the strengths
significantly overweigh the weaknesses. The calculation of late-time,
nebular spectra involves many uncertainties, and hence I do not regard
weakness~(3) above as fatal (although clearly more work will be required
to explain it away). Both weaknesses~(1) and~(2) can be overcome if it
can be demonstrated that SNe~Ia statistics can be reproduced within the
uncertainties that still plague the theoretical population synthesis
models. As I will show in \S5, this is indeed the case.

\subsubsection{Sub-Chandrasekhar mass helium ignitors}

In these models a C--O WD accumulates a helium layer of
$\sim0.15$~M$_{\odot}$ with a total mass that is sub-Chandrasekhar. The
helium ignites off-center (at the bottom of the layer), resulting in an
event known as ``Indirect Double Detonation'' (IDD) or ``Edge Lit
Detonation'' (ELD). Basically, one detonation propagates outward
(through the helium), while an inward propagating pressure wave
compresses the C--O core which ignites off-center, followed by an
outward detonation (e.g.\ Livne 1990; Livne \& Glasner 1991; Woosley \&
Weaver 1994; Livne \& Arnett 1995; Hoeflich \& Khokhlov 1996; and 
Ruiz-Lapuente, talk presented at the Chicago meeting on Type~Ia 
Supernovae: Theory and Cosmology, October 1998).

The main {\it strengths\/} of ELD (sub-Chandrasekhar) models are:
\begin{enumerate}
\item[(1)] It is easier to achieve the required {\it statistics\/},
since less mass needs to be accreted, and the WD does not need to be
extremely massive (e.g.\ Ruiz-Lapuente, Canal \& Burkert 1997;
Di~Stefano et~al.\ 1997; Yungelson \& Livio 1998).
\item[(2)] The {\it late-time spectrum\/} (in particular the Fe~III
feature at $\sim4700$~\AA) agrees better with ELD models.
\item[(3)] SNe~Ia {\it light curves\/} can be reproduced adequately by
ELD models (although the light curves rise somewhat faster than
observed, due to $^{56}$Ni heating; Hoeflich et~al.\ 1997).
\end{enumerate}

The main {\it weaknesses\/} of ELD models are:
\begin{enumerate}
\item[(1)] The {\it spectra\/} that are produced by ELD models generally
do not agree with observations (e.g.\ of SN~1994D; Nugent et~al.\ 1997).
The agreement is somewhat better for the subluminous SNe~Ia (e.g.\
SN~1991bg; Nugent et~al.\ 1997; Ruiz-Lapuente, talk presented at the 
Chicago meeting on Supernovae, October 1998), but even
there it is not very good.
\item[(2)] The {\it highest velocity ejecta have the wrong
composition\/} ($^{56}$Ni and He; not intermediate mass elements; also 
no high velocity~C; e.g.\ Livne \&
Arnett 1995). This is due to the fact that in these models, essentially
by necessity, the intermediate mass elements are sandwiched by Ni and
He/Ni rich layers.
\item[(3)] Since ELD models allow for a range of WD masses, and since
more massive WDs produce brighter SNe, one might expect this model to
produce a more gradual decline on the bright side of the {\it luminosity
function\/}, in contradiction to the observed sharp decline (see \S2
characteristic~(3)).
\end{enumerate}

My overall assessment of the sub-Chandrasekhar mass model is that the
weaknesses (and in particular weakness~(2) which appears almost
inevitable) greatly overweigh the strengths in terms of this being a
model for the bulk of SNe~Ia. It is still possible that ELDs may
correctly represent some subluminous SNe~Ia (e.g.\ Ruiz-Lapuente, Canal,
\& 
Burkert 1997). I should note that Pinto (verbal communication at the
Chicago
meeting on Supernovae) insists that his ELD models manage to 
overcome all of the above weaknesses and that they are able to produce 
excellent fits to both light curves and spectra. By the time of the
writing 
of this review, however, I have unfortunately failed to find published
results 
of these models and hence I cannot comment on them.

\subsection{The favored model}

On the basis of the above discussion the basic model can be further
refined, and I tentatively conclude that: {\it SNe~Ia represent
thermonuclear disruptions of mass accreting C--O white dwarfs, when
these white dwarfs reach the Chandrasekhar limit and ignite carbon at
their centers\/}!

\section{The two possible scenarios}

The next step in the search for the progenitor systems of SNe~Ia is even
more
controversial. Two possible scenarios have been proposed: (i)~The {\it
double-degenerate\/} scenario, in which two CO WDs in a binary system
are brought together by the emission of gravitational radiation and
coalesce (Webbink 1984; Iben \& Tutukov 1984). (ii)~The {\it 
single-degenerate\/} scenario, in which a CO WD accretes hydrogen-rich
or helium-rich material from a non-degenerate companion (Whelan \& Iben
1973; Nomoto 1982).

In the first scenario the progenitor systems are necessarily {\it binary
WD systems\/} in which the total mass exceeds the Chandrasekhar mass,
and which have binary periods shorter than about thirteen hours (to
allow merger within a Hubble time).

In the second scenario the progenitors could be systems like: (i)~{\it
Recurrent novae\/} (both of the type in which the WD accretes hydrogen
from a giant like T~CrB, RS~Oph and of the type in which the WD accretes
helium rich material from a subgiant like U~Sco, V394~CrA, and Nova
LMC~1990\#2), (ii)~{\it Symbiotic Systems\/} (in which the WD accretes
from a low mass red giant), or (iii)~persistent {\it Supersoft X-ray
Sources\/} (in which the WD accretes at a high rate 
$\apgt10^{-7}$~M$_{\odot}$/yr from a subgiant companion).

I will now examine the strengths and weaknesses of each one of these
scenarios.

\subsection{The double-degenerate scenario}

There is no question that binary white dwarf systems are an expected
outcome of binary star evolution (e.g.\ Iben \& Tutukov 1984; Iben \&
Livio 1993). Once the lighter WD (which has a larger radius) fills its
Roche lobe, it is entirely dissipated within a few orbital periods, to
form a massive disk around the primary (e.g.\ Rasio \& Shapiro 1994).
The subsequent evolution of the system depends largely on the accretion
rate through this disk (e.g.\ Mochkovitch \& Livio 1990; see discussion 
below).

The main {\it strengths\/} of this scenario are the following:
\begin{enumerate}
\item[(1)] The {\it absence of hydrogen\/} in the spectrum is naturally
explained in a model which involves the merger of two C--O WDs. In fact,
if hydrogen is ever detected in the spectrum of a SN~Ia, this would deal
a fatal blow to this model. Tentative evidence for circumstellar
H$\alpha$ absorption is SN~1990M was presented by Polcaro and Viotti
(1991). However, Della Valle, Benetti \& Panagia (1996) demonstrated
convincingly that the absorption was caused by the parent galaxy, rather
than by the SN environment.
\item[(2)] In spite of some impressions to the contrary, {\it many
double WD systems do exist\/}. In a sample of 153 field WDs and subdwarf
B~stars, Saffer, Livio \& Yungelson (1998) found 18 new 
double-degenerate candidates. There are currently eight known systems
with
orbital periods of less than half a day. While only one of those systems
(KPD~0422+5421; Koen, Orosz \& Wade (1998)) has a total mass which
within the
errors could be higher than the Chandrasekhar mass, the sample of
confirmed short-period double-degenerates is still smaller than the
number predicted to contain a massive system.
\item[(3)] Population synthesis calculations predict the {\it right
statistics\/} for mergers, about $10^{-3}$~yr$^{-1}$ events for
populations that are $\sim10^8$~yr old and $10^{-4}$~yr$^{-1}$ for
populations that are $\sim10^{10}$~yr old.
\item[(4)] Since double WD systems were found to exist, {\it mergers\/}
with some ``interesting'' consequences (either a SN~Ia or an 
accretion-induced collapse) appear inevitable.
\item[(5)] The explosion or collapse is expected to occur at the
Chrandrasekhar mass, which as I noted in \S4.3, I regard as a property
of the favored model.
\end{enumerate}

The main {\it weaknesses\/} of the double-degenerate scenario are the
following:
\begin{enumerate}
\item[(1)] There are strong indications that WD mergers may lead to 
off-center carbon ignition, accompanied by the conversion of the C--O WD
to an O--Ne--Mg composition, and followed by an accretion-induced
collapse rather than a SN~Ia (e.g.\ Mochkovitch \& Livio 1990; Saio \&
Nomoto 1985, 1998; Woosley \& Weaver 1986). 
\item[(2)] Galactic chemical evolution results, and in particular the
behavior of the [O/Fe] ratio as a function of metallicity ([Fe/H]) have
been claimed to be inconsistent with WD mergers as the mechanism for
SNe~Ia (Kobayashi et~al.\ 1998).
\end{enumerate}

Since we are now getting to the final stages in the identification of
the progenitors, it is important to assess critically the severity of
the above weaknesses. I will therefore discuss now each one of them in
some detail.

\subsubsection{Constraints from Galactic chemical evolution}

Supernovae Type~II  (SNe~II) are explosions resulting from the core
collapse of massive ($\apgt8$~M$_{\odot}$) stars. These supernovae
produce relatively more oxygen and magnetism than iron ([O/Fe]~$>0$). On
the other hand SNe~Ia produce mostly iron and little oxygen. Until
recently, the impression has been that metal poor stars
([Fe/H]~$\leq-1$) 
have a nearly flat relation of [O/Fe] vs.\ [Fe/H], with a value of
[O/Fe]~$\sim0.45$ (e.g.\ Nissen et~al.\ 1994), while disk stars
([Fe/H]$\apgt-1$) show a linearly decreasing [O/Fe] with increasing
metallicity (e.g.\ Edvardsson et~al.\ 1993). The break near
[Fe/H]~$\sim-1$ was traditionally explained by the fact that the early
heavy element production was done exclusively by SNe~Ia, with the break
occurring when the larger Fe production by SNe~Ia kicks in (e.g.\
Matteucci \& Greggio 1986).

Recently, Kobayashi et~al.\ (1998) performed chemical evolution
calculations for both the double-degenerate scenario and for the 
single-degenerate scenario. For the latter they used two types of
progenitor systems: one with a red giant companion and an orbital period
of tens to hundreds of days, and the other with a near main sequence
companion and a period of a few tenths of a day to a few days.

They obtained for the double-degenerate scenario (for which they took a
time delay of $\sim0.1$--0.3~Gyr) a break at [Fe/H]~$\sim-2$. For the
single-degenerate scenario (with a delay caused by the main sequence
lifetime of $\apgt1$~Gyr; including metallicity effects), they obtained
a break at [Fe/H]~$\sim-1$.  Kobayoshi et~al.\ (1998) thus concluded
that
the double-degenerate scenario is inconsistent with Galactic chemical
evolution results.

Personally, I am not convinced by this apparent discrepancy, since
Galactic chemical evolution calculations (and observations) are
notoriously uncertain. In particular, the most recent Keck observations
of oxygen in unevolved metal-poor stars show {\it no break\/} in the
[O/Fe] vs.\ [Fe/H] relation. Rather, oxygen is enhanced relative to iron
over three orders of magnitude in [Fe/H] in a robustly linear relation
(Boesgaard et~al.\ 1999). Consequently, apparent inconsistencies based
on Galactic chemical evolution cannot be regarded at present as a
fatal weakness of the double-degenerate scenario.

\subsubsection{SN~Ia or accretion induced collapse?}

Potentially a more serious (and possibly even fatal) weakness of the
double-degenerate scenario comes from the fact that some estimates and
calculations indicate that the coalescence of two C--O WDs may lead to
an accretion-induced collapse rather than to a SN explosion (e.g.\
Mochkovitch \& Livio 1990; Saio \& Nomoto 1985, 1998; Kawai, Saio \&
Nomoto 1987; Timmes, Woosley \& Taam 1994).

The point is the following: once the lighter WD fills its Roche lobe, it
is dissipated within a few orbital periods (Benz et~al.\ 1990; Rasio \&
Shapiro 1995; Guerro 1994) and it forms a hot thick disk configuration
around the more massive white dwarf. This disk is mainly rotationally
supported and hence central carbon ignition does not take place
immediately, but rather the subsequent evolution depends largely on the
rate of angular momentum transport and removal, since they determine the
accretion rate onto the primary WD. As long as the accretion rate is
higher than about $\dot{M}\apgt2.7\times10^{-6}$~M$_{\odot}$ yr$^{-1}$,
{\it carbon is ignited off-center\/} (at the core-disk boundary; this
may happen during the merger itself; e.g.\ Segretain  1994). Under such
conditions, the flame was found (in spherically symmetric calculations)
to propagate all the way to the center within a few thousand years, thus
burning the C--O into an O--Ne--Mg mixture with {\it no explosion\/}
(i.e.\ before carbon is centrally ignited; e.g.\ Saio \& Nomoto 1998).
Such configurations are expected to collapse (following electron
captures on $^{24}$Mg) to form neutron stars (Nomoto \& Kondo 1991;
Canal
1997). The main questions are then:
\begin{enumerate}
\item[(i)] What accretion rates can be expected from the initial 
WD-thick disk configuration? 
\item[(ii)] May some aspects of the flame propagation be different given
the fact that the real problem is three-dimensional while most of the
existing calculations were performed using a spherically symmetric code?
In particular, {\it could the carbon burning be quenched before the
transformation to O--Ne--Mg composition occurs\/}?
\end{enumerate}

The answers to both of these questions involve uncertainties, however
some possibilities are more likely than others.  First, it appears {\it
very difficult to avoid high accretion rates\/}. If the MHD turbulence
that is expected to develop in accretion disks (e.g.\ Balbus \& Hawley
1998) is operative, with a corresponding viscosity parameter of
$\alpha\sim0.01$ (where the viscosity is given by $\nu\sim\alpha c_{\rm
s}H$, with $H$ being a vertical scaleheight in the disk and $c_{\rm s}$
the speed of sound; e.g.\ Balbus, Hawley \& Stone 1996), then angular
momentum can be removed in a matter of days! In such a case, even if the 
accretion rate is Eddington limited (at $\sim10^{-5}$~M$_{\odot}$/yr),
off-center carbon ignition should still occur, with an eventual collapse
rather than an explosion. Deviations from spherical symmetry can only
hurt, since they may allow accretion to proceed at a super-Eddington
rate.
It is difficult to see why the dynamo-generated viscosity would be
suppressed for the kind of shear and temperatures expected in the disk.

Concerning the burning itself, recent attempts at multi-dimensional
calculations of the flame propagation and a more detailed analysis of
some of the processes involved (Garcia-Senz, Bravo \& Serichol 1998;
Bravo \& Garcia-Senz 1999) indicate that if anything, accretion induced
collapses are an even more likely outcome than previously thought. This
is due to the effects of electron captures in Nuclear Statistical
Equilibrium which tend to stabilize the thermonuclear flame, and to
Coulomb corrections to the equation of state. The latter has the effect
of reducing the flame velocities and the electronic and ionic pressures,
all of which result in a reduction in the critical density which
separates explosions from collapses.

Finally, on the observational side there are also two points which argue
at some level against WD mergers as SNe~Ia progenitors.
\begin{enumerate}
\item[(i)] Even if MHD viscosity could somehow be suppressed, and the
disk surrounding the primary WD could cool down, so that angular 
momentum would be transported only via the viscosity of (partially) 
degenerate electrons, this would result
in an accretion timescale of $\sim10^9$~yrs (Mochkovitch \& Livio 1990;
Mochkovitch et~al.\ 1997). The system prior to the explosion would have
an absolute magnitude of $M_{\rm V}\aplt10$ (with much of the emission
occurring in the UV). There is no evidence for the existence of some
$\sim10^7$ such objects in the Galaxy. 
\item[(ii)] The existence of planets around the pulsars PSR~1257+12 and
PSR~1620--26 (Wolszczan 1997; Backer 1993; Thorsett, Arzoumanian \&
Taylor 1993) could be taken to mean (this is a model dependent
statement) that mergers tend to produce accretion induced collapses
rather 
than SNe~Ia. In one of the leading models for the formation of such
planets (Podsiadlowski; Pringle \& Rees 1991; Livio, Pringle \& Saffer
1992), the planets form in the following sequence of events. The lighter
WD is dissipated (upon Roche lobe overflow) to form a disk around the
primary. As material from this disk is accreted, matter at the outer
edge of the disk has to absorb the angular momentum, thereby expanding
the disk to a large radius. The planets form from this disk in the
some way that they did in the solar system, while the central object
collapses to form a neutron star.
\end{enumerate}

\subsubsection{Overall assessment of the double-degenerate scenario}

It has now been observationally demonstrated that many double-degenerate
systems exist. The general agreement between the distribution of the
observed properties (e.g.\ orbital periods, masses) and those predicted
by population synthesis calculations (Saffer, Livio \& Yungelson 1998), 
suggests that
the fact that no clear candidate (short period) system with a total mass
exceeding the Chandrasekhar mass has been found yet, may merely reflect
the insufficient size of the observational sample. Thus, there is very
little doubt in my mind that statistics is not a serious problem. The
most disturbing uncertainty is related to the outcome of the merger
process itself. The discussion in \S5.1.2 suggests that {\it collapse to
a
neutron star is more likely than a SN~Ia\/} (see also Mochovitch et~al.\
1997).

\subsection{The single-degenerate scenario}

The main {\it strengths\/} of the single degenerate scenario are:
\begin{enumerate}
\item[(1)] A class of objects in which hydrogen is being transferred at
such high rates that it {\it burns steadily\/} on the surface of the WD
has been identified---the Supersoft X-ray Sources (Greiner, Hasinger \& 
Kahabka 1991; van den Heuvel et~al.\ 1992; Southwell et~al.\ 1996;
Kahabka 
and van den Heuvel 1997). If the accreted matter can indeed be retained, 
this provides a natural path to an increase in the WD mass towards the 
Chandrasekhar mass (e.g.\ Di~Stefano \& Rappaport 1994; Livio 1995,
1996a; 
Yungelson et~al.\ 1996).
\item[(2)] Other candidate progenitor systems are known to exist, like
symbiotic systems (e.g.\ Munari \& Renzini 1992; Kenyon et~al.\ 1993;
Hachisu, Kato \& Nomoto 1999) and recurrent novae (Hachisu et~al. 1999).
\item[(3)] There have been claims that the single degenerate scenario
fits better the results of Galactic chemical evolution (e.g.\ Kobayoski
et~al.\ 1998). However, as I have shown in \S5.1.1, recent observations
cast doubt on this assertion. Similarly, nucleosynthesis results show
that in order to avoid unacceptably large ratios of $^{54}$Cr/$^{56}$Fe 
and $^{50}$Ti/$^{56}$Fe, the central density of the WD at the moment of
thermonuclear runaway must be lower than $\sim2\times10^9$~g cm$^{-3}$
(Nomoto et~al.\ 1997). Such low densities are realized for high
accretion rates ($\apgt10^{-7}$~M$_{\odot}$ yr$^{-1}$), which are
typical for the Supersoft X-ray Sources. Nucleosynthesis results suffer
too, however, from considerable uncertainties (e.g.\ Nagataki, Hashimoto
\& Sato 1998).
\end{enumerate}

The main {\it weaknesses\/} of the single degenerate scenario are:
\begin{enumerate}
\item[(1)] The upper limits on {\it radio detection\/} of hydrogen at
2~and 6~cm
in SN~1986G, taken approximately one week before optical maximum (Eck
et~al.\ 1995), rule out a symbiotic system progenitor for this system
with a wind mass loss rate of $10^{-7}\aplt \dot{M}_{\rm W}
\aplt10^{-6}$~M$_{\odot}$ yr$^{-1}$ (Boffi \& Branch 1995). This in
itself is not fatal, since SN~1986G is somewhat peculiar (e.g.\ Branch
and van den Bergh 1993), and the upper limit on the mass loss rate is at
the high end of observed symbiotic winds. An even less stringent upper
limit from x-ray and H$\alpha$ observations exists for SN~1994D (Cumming
et~al.\ 1996).
\item[(2)] There exists some uncertainty whether WDs can then reach the
Chandrasekhar mass {\it at all\/} by the accretion of hydrogen (e.g.\
Cassisi, Iben \& Tornambe 1998). Furthermore, even if they can, the
question of whether they can produce the required SNe~Ia statistics is
highly controversial (e.g.\ Yungelson et~al.\ 1995, 1996; Yungelson \&
Livio 1998; Hachisu, Kato \& Nomoto 1999; Hachisu et~al.\ 1999).
\end{enumerate}

I will now examine these weaknesses in some detail.

\subsubsection{Observational detection of hydrogen}

Ultimately, the presence or total absence of hydrogen in SNe~Ia will
distinguish unambiguously between single-degenerate and 
double-degenerate models. To date, hydrogen has not been convincingly
detected
in {\it any\/} SN~Ia. It is interesting to note that narrow
$\lambda6300$, $\lambda6363$~[OI] lines were observed only in one SN~Ia
(SN~1937C; Minkowski 1939), but even in that case there was no hint of a
narrow H$\alpha$ line. Hachisu, Kato \& Nomoto (1999) estimate in one of
their models (which involves stripping of material from the red giant;
see below) a density measure of $\dot{M}/v_{10}\sim10^{-8}$~M$_{\odot}$
yr$^{-1}$ (where $v_{10}$ is the wind velocity in units of 
10~km~s$^{-1}$), while the most stringent radio upper limit existing
currently (for SN~1986G) is $\dot{M}/v_{10}\sim10^{-7}$~M$_{\odot}$ 
yr$^{-1}$ (Eck et~al.\ 1995; for SN~1994D Cumming et~al.\ (1996) find
from
H$\alpha$ an upper limit of $\dot{M}\sim1.5\times
10^{-5}$~M$_{\odot}$ yr$^{-1}$ for a wind speed of 10~km~s$^{-1}$; for
SN~1992A Schlegel \& Petre (1993) find from X-ray observations an upper 
limit of $\dot{M}/v_{10}=(2-3)\times10^{-6}$~M$_{\odot}$ yr$^{-1}$).
Thus, 
while it is impossible at present to rule out single-degenerate models
on 
the basis of the apparent absence of hydrogen, the hope is that near 
future observations will be able to determine definitively whether this 
absence is real or if it merely represents the limitations of existing 
observations (an improvement by two orders of magnitude will give a 
definitive answer).

\subsubsection{Statistics}

Growing the WD to the Chandrasekhar mass is not easy. At accretion rates
below $\sim10^{-8}$~M$_{\odot}$/yr WDs undergo repeated nova outbursts
(e.g.\ Prialnik \& Kovetz 1995), in which the WDs lose more mass than
they accrete between outbursts (e.g.\ Livio \& Truran 1992). For
accretion rates in the range $10^{-8}$--a few
$\times10^{-7}$~M$_{\odot}$/yr,
while helium can accumulate, the WDs experience mass loss due to helium
shell flashes and due to the common envelope phase which results from
the engulfing of the secondary star in the expanding envelope (with mass
loss
occurring due to drag energy deposition). At accretion rates above a few 
$\times10^{-7}$~M$_{\odot}$/yr, the WDs expand to red giant
configurations 
and lose mass due to drag in the common envelope and due to winds (e.g.\ 
Cassisi et~al.\ 1998). The net result of this has been that population 
synthesis calculations which follow the evolution of all the binary
systems 
in the Galaxy, tended until recently to conclude that single degenerate
channels manage to bring WDs to the Chandrasekhar mass only at about
10\% of the inferred SNe~Ia frequency of $4\times10^{-3}$~yr$^{-1}$
(e.g.\ 
Yungelson et~al.\ 1995, 1996; Yungelson \& Livio 1998; Di~Stefano
et~al.\ 
1997; although see Li \& van den Heuvel 1997).

Very recently, a few serious attempts have been made to investigate
whether
the statistics could be improved by increasing the ``phase space'' for
single degenerate scenarios, given the fact that population synthesis
calculations involve many assumptions. These attempts resulted in three
directions in which the phase space could be increased. 
\begin{enumerate}
\item[(i)] The accumulation efficiency of helium has been recalculated
using OPAL opacities (Kato \& Hachisu 1999). These authors concluded
that helium can accumulate much more efficiently than found by Cassisi
et~al.\ (1998), mainly because the latter authors used relatively low
WD masses (0.516~M$_{\odot}$ and 0.8~M$_{\odot}$) and old opacities in
their calculations.
\item[(ii)] Hachisu et~al.\ (1999) claimed to have identified an
evolutionary channel for single-degenerate systems previously overlooked
in population synthesis calculations. In this channel, the C--O WD is
formed from a red giant with a helium core of 0.8--2.0~M$_{\odot}$
(rather than from an asymptotic giant branch star with a C--O core). The
immediate progenitors in this case are expected to be either helium-rich
Supersoft X-ray Sources or recurrent novae of the U~Sco subclass.
\newpage
\item[(iii)] It has been suggested that the inclusion of a few
additional physical effects, can increase substantially the phase space 
of the symbiotic channel (Hashisu, Kato \& Nomoto 1999). These new 
effects included:
\end{enumerate}
\begin{enumerate}
\item[(1)] The WD loses much of the transferred mass in a massive wind.
This has the effect that the mass transfer process is stabilized for a
wider range of mass ratios, up to $q_{\rm max}\equiv m_2/m_1=1.15$
instead 
of $q_{\rm max}=0.79$ without the massive wind.
\item[(2)] It has been suggested that the wind from the WD strips the
outer layers of the red giant at a high rate. This increases the allowed
mass ratios (for stability) even above 1.15, essentially indefinitely.
\item[(3)] It has been suggested that at large separations (up to
$\sim30,000$~R$_{\odot}$) the wind from the red giant acts like a common
envelope to reduce the separation, thus allowing much wider initial 
separations to result in interaction.
\end{enumerate}

There are many uncertainties associated with all of these attempts to
increase the phase space. For example, the efficiency of mass stripping
from the giant by the wind from the WD may be much smaller than assumed
by Hachisu et~al.\ (1999), for the following reasons. At high accretion
rates, much of the mass loss from the WD may be in the form of an
outflow or a collimated jet, perpendicular to the accretion disk rather
than in the direction of the giant. Evidence that this is the case is
provided by the jet satellite lines to He~II~4686, H$\beta$ and
H$\alpha$
observed in the Supersoft X-ray Source RX~J0513.9$-$6951 (Southwell
et~al.\
1996). These jet lines are very similar to those seen in the
prototypical jet source SS~433 (e.g.\ Vermeulen et~al.\ 1992).
Furthermore, even if some of the WD wind hits the surface of the giant,
it is not clear how efficient it would be in stripping mass, since the
rate of energy deposition per unit area by the wind is smaller by two
orders of magnitude that the giant's own intrinsic flux.

Similarly, the efficiency of helium accumulation is still highly
uncertain, as the differences between the results of Kato \& Hachisu
(1999) and Cassisi et~al.\ (1998) have shown. 

Finally, all the new suggestions for the increase in phase space rely
very heavily on the results of the wind solutions of Kato (1990; 1991),
which involve a treatment of the radiation not nearly as sophisticated
as that of more state of the art radiative transfer codes (e.g.\
Hauschildt et~al.\ 1995, 1996).

\subsubsection{Overall assessment of the single-degenerate scenario}

The above discussion suggests that probably not all the scenarios for
increasing the ``phase space'' of the single-degenerate channels work
(if they did we might have had the opposite problem of too high a
frequency of SNe~Ia!).  However, these attempts serve to demonstrate
that the input physics to population synthesis codes still involves many
uncertainties. My feeling is therefore that given the many potential
channels leading to SNe~Ia, statistics should not be regarded as a
serious problem.

Single-degenerate scenarios therefore appear quite promising, since
unlike the situation a decade ago, a class of objects in which the WDs
accrete hydrogen steadily (the Supersoft X-Ray Sources) has actually
been identified. The main problem with single-degenerate scenarios
remains the non-detection of hydrogen so far. While a difficult
observational problem (see \S6), the establishment of the presence or
absence of hydrogen in SNe~Ia should become a first priority for SNe
observers.

\subsection{What if....?}

Given the fact that there are still uncertainties involved in
identifying the SNe~Ia progenitors, and that WD mergers and some 
form of off-center helium ignitions 
 almost certainly occur, it is instructive to pose a few
``what if'' questions. For example:  {\it What if WD mergers with a
total mass exceeding Chandrasekhar do not produce SNe~Ia, what do they
produce then\/}? The answer in this case will have to be that they
almost certainly produce either neutron stars via accretion induced
collapses, or single WDs, if the merger is accompanied by extensive mass
loss from the system.

{\it What if off-center helium ignitions do not produce SNe~Ia\/}? In
this 
case, if an explosive event indeed ensues, a population of ``super
novae'' 
(with $\sim0.15$~M$_{\odot}$ of $^{56}$Ni and He) is yet to be detected 
(maybe SN~1885A in M31 was such an event?).  {\it What if off-center
helium
ignitions do produce SNe~Ia? What comes out of the systems with $M_{\rm
WD}\apgt1$~M$_{\odot}$, which should be even brighter\/}? It is
difficult to believe that the latter are represented by the very few
bright objects like SN~1991T. Thus, we see that off-center helium
ignitions
seem to present an observational problem both if they {\it do\/} and if
they 
{\it do not\/} produce SNe~Ia. To me this suggests that the physics of
these 
events is not well understood (for example, maybe off-center helium
ignition 
fails to ignite the C--O core after all).

\section{How can we hope to identify the progenitors?}

There are several ways in which observations of {\it nearby\/}
supernovae could
solve the mystery of SNe~Ia progenitors:
\begin{enumerate}
\item[(1)] A combination of {\it early high resolution optical
spectroscopy, x-ray observations\/} and {\it radio observations\/} can
both provide limits on $\dot{M}/v$ from the progenitors and potentially
detect the presence of circumstellar hydrogen (if it exists).

\item []For example, narrow HI in emission or absorption could be
detected
either very early, or shortly after the ejecta become optically thin
($\sim100$~days). The latter is true because the SN ejecta probably
engulfs the companion at early times (e.g.\ Chugai 1986; Livne, Tuchman
\& Wheeler 1992). The interaction of the ejecta with the circumstellar
medium can be observed either in the radio (e.g.\ Boffi \& Branch 1995)
or in x-rays (e.g.\ Schlegel 1995). The collision of the ejecta (with
circumstellar matter) can also set up a forward and a reverse shock
(e.g.\ Chevalier 1984; Fransson, Lundqvist \& Chevalier 1996), and
radiation from the latter can ionize the wind and produce H$\alpha$ 
emission (e.g.\ Cumming et~al.\ 1996).
\item[(2)] Early observations of the gamma-ray light curve (or gamma-ray
line profiles) could distinguish between carbon ignitors and 
sub-Chandrasekhar helium ignitor models (see \S4.2.2) since the  latter 
can be expected to result in a quicker rise of the gamma-ray light curve 
due to the presence of $^{56}$Ni in the outer layers (and different 
gamma-ray line profiles; because of the high velocity $^{56}$Ni).
\item[] Observations of very distant supernovae (at $z\sim3$--4) with
the Next Generation Space Telescope (NGST) can also help (e.g.\
Yungelson \& Livio 1999). For example, the progenitors can be identified 
from the observed frequency of SNe~Ia as a function of redshift (e.g.\
Yungelson \& Livio 1998, 1999; Ruiz-Lapuente \& Canal 1998; Madau, Della
Valle \& Panagia 1998; and see \S8), since different progenitor models
produce different redshift distributions. Personally, I think that it
would be absolutely pathetic to have to resort to this possibility.
Rather, one would like to identify the progenitors independently, and to
use the observations of supernovae at high~$z$ to constrain models of
cosmic evolution of rates, luminosity, and input into galaxies.
\end{enumerate}

\section{Could we be fooled?}

One of the key questions that result from the fact we do not know with
certainty which systems are the progenitors of SNe~Ia is clearly: {\it
is it possible that SNe~Ia at higher redshifts are systematically dimmer
than their low-redshift counterparts\/}? In this respect it is important
to remember that a systematic decrease in the brightness by $\sim0.25$
magnitudes is sufficient to explain away the need for a cosmological
constant. In a recent work, Yungelson \& Livio (1999) calculated the
expected ratio of the rate of SNe~Ia to SNe from massive stars
(Types~II, Ia, Ic) as a function of redshift for several progenitor
models. They showed that if different progenitor systems can contribute
to the total SNe~Ia rate (e.g.\ double-degenerates at the Chandresekhar
mass and single-degenerates with subgiant donors at sub-Chandrasekhar
masses), then it is possible in principle that a different progenitor
class will start to dominate at $z\sim1$. However, such a transition is
highly unlikely, because: (i)~in some models the transition has the
opposite effect to the observed one (e.g.\ double degenerates which may
be expected to be brighter than sub-Chandrasekhar ELDs dominate at the
higher redshifts). (ii)~If the contribution from physically different 
channels was indeed significant, one would have expected to observe 
this division more clearly also in the local sample, which is not the 
case ($\sim90$\% of SNe~Ia are ``normals''). Consequently, I do not 
believe that the observed universal acceleration is an artifact of 
the observed sample being dominated by different progenitor classes.

\section{Tentative conclusions and observational tests}

On the basis of the analysis and discussion in the present work, the
following tentative conclusions can be drawn:
\begin{enumerate}
\item[(1)] SNe~Ia are almost certainly thermonuclear disruptions of mass
accreting {\it C--O white dwarfs\/}.
\item[(2)] It is very likely that the explosion occurs {\it at the
Chandrasekhar mass\/}, as {\it carbon is ignited at the WD center\/}.
Off-center ignition of helium at sub-Chandrasekhar masses may still be
responsible for a subset of the SNe~Ia which are subluminous, but this
is less clear.
\item[(3)] The immediate progenitor systems are still not known with
certainty. From the discussion in \S5 (see in particular \S5.1.3 and
5.2.3) however, I conclude that presently {\it single degenerate
scenarios look more promising\/}, with hydrogen or helium rich material
being transferred from a subgiant or giant companion (systems like
Supersoft X-Ray Sources and Symbiotics).
\item[(4)] Definitive answers concerning the nature of the progenitors
can be obtained from observations taken as early as possible in: {\it 
x-rays, radio, and high resolution optical spectroscopy. The
establishment of the presence or absence of hydrogen in SNe~Ia should be
regarded as an extremely high priority goal for supernovae observers\/}.
If hydrogen will not be detected at interesting limits (corresponding to
$\dot{M}/v_{10}\sim10^{-8}$~M$_{\odot}$ yr$^{-1}$), this will point
clearly towards the double-degenerate scenario.
\item[(5)] Observations of SNe~Ia at high redshifts can help to test
particular ingredients of the models which are directly related to the
nature of the progenitors. For example, most of the models aiming at
improving the statistics of the single-degenerate scenarios rely on a
strong wind from the accreting WD. These models thus predict an
``inhibition'' of SNe~Ia in low-metallicity environments, and in
particular a significant decrease in the rate of SNe~Ia at $z\sim1$--2
(Kobayashi et~al.\ 1998). At present, the detection of a very likely
SN~Ia at redshift $z=1.32$ (SN~1997ff) in the Hubble Deep Field appears 
inconsistent with this prediction (Gilliland, Nugent \& Philips 1999), 
but more observations will be required to give a more definitive answer.
\end{enumerate}

\begin{acknowledgments}
This research has been supported in part by NASA Grant NAG5--6857.
\end{acknowledgments}

\end{document}